\documentstyle{mn}
\input{epsf}

\title[`Negative' intensity patches]
{`Negative' intensity patches in angular variations 
of CMB as a probe of the period of reionization}

\author[Doroshkevich \&  Dubrovich]
       {A. Doroshkevich$^{1,2}$, \&  V. Dubrovich$^{1,3}$\\
	$1$Theoretical Astrophysics Center,
          Juliane Maries Vej 30,
          DK-2100 Copenhagen \O, Denmark\\
	$2$Keldysh Institute of Applied Mathematics,
                        Russian Academy of Sciences,
                        125047 Moscow,  Russia\\
	$3$ SAO, 196140, Pulkovo, St.-Petersburg, Russia, \\
}
\date{Accepted ...,
      Received ;
	in original form }

\begin{document}
\maketitle

\begin{abstract}
The observational tests for the period of reionization of the 
universe are discussed. We show that this period can be observed 
as {\it negative} intensity patches of the CMB radiation with 
the amplitude $\delta T/T\sim 10^{-5}$ and the angular sizes 
$\theta_T\sim$10 angular seconds in range of the wavelength
0.1 cm$\leq\lambda\leq$1 cm. The expected number density and 
frequency dependence of the amplitude permit to recognize this 
effect and to discriminate it from the noise. This method applied 
to the small scale variations of CMB temperature complements well 
the traditional spectral approach.

The number density and the amplitude of observed 'negative' 
intensity patches depend upon the redshift of reionization that 
allow to estimate roughly this redshift. The ionized bubbles 
formed just before the period of reionization could also be 
seen as the highest peaks. The expected results are sensitive 
to the Jeans scale at the period of reionization, to the small 
scale shape of the primordial power spectrum and to the mass of 
dark matter particles. 
\end{abstract}


\section{Introduction}

The analysis of the cosmic microwave background radiation 
(CMB) anisotropy generated after reionization of the 
universe has been began in Sunyaev \& Zel'dovich (1980) 
(static and kinematic Sunyaev-Zel'dovich effect), Kaiser 
(1984) and Ostriker \& Vishniac (1986). Later on various 
aspects of this problem were discussed in many publications 
(see, e.g., Tegmark, Silk, Blanchard, 1994; Hu, Scott \& Silk
1994; Persi 1995; Tegmark \& Silk 1995; Aghanim et al. 1996; 
Hu \& White (1996); Knox, Scoccimarro \& Dodelson 1998; Gruzinov 
\& Hu 1998; Peebles \& Juszkiewicz 1998, Griffiths, Barbosa \& 
Liddle 1999; Haiman \& Knox 1999; Peterson et al. 1999; 
Weller, Battye \& Albrecht 1999; Rees 1999;
Molnar \& Birkinshaw 2000; Cooray, Hu, \& Tegmark 2000; Benson et 
al. 2001). Various possible sources of such perturbations were 
considered -- from linear and second order perturbations and 
up to CMB scattering within a set of high density clouds and 
clusters of galaxies.   

Firstly the expected variations of the CMB temperature have 
been calculated in Vishniac (1987) where has been noted that, for 
larger $l$, the power spectrum of the temperature variations 
reproduces well the spectrum of perturbations and is defined 
by the reionization epoch. Recent publications repeat 
the same calculations, concentrate more attention on the impact 
of nonlinearity and inhomogeneity of reionization (Peebles \& 
Juszkiewicz 1998; Gruzinov \& Hu 1998; Knox, Scoccimarro \& 
Dodelson 1998; Jaffe \& Kamionkowski 1998; Haiman \& Knox 1999; 
Hu 2000), and use numerical simulations of the 
reionization (Springel et al. 2000;  Bruscoli, Ferrara \& Ciardi 
2000; Gnedin and Jaffe 2000) for direct calculations of the 
power spectrum of CMB variations, $C_l$. It is interesting that 
results based on simulations differ in some respects from 
theoretical expectations. Thus, for example, Peebles \& 
Juszkiewicz (1998) demonstrate the small impact of moderate 
nonlinearity which can be expected during the period of 
reionization. The toy model discussed in Gruzinov \& Hu (1998) 
shown that the ionization in patches also cannot essentially 
distort expectations obtained under assumption of homogeneous 
reionization. In simulations, however, the noticeable impact 
of both effects was found. The more detailed discussions and 
comparisons of these results with theoretical expectations are 
needed to clarify the validity of theoretical assumptions and 
representativity of simulations. 

Here we turn back to the simplest case of linear theory 
and consider the small scale angular variations of the 
CMB temperature produced by velocity perturbations after 
the reionization of the universe (linear Doppler and Ostriker-
Vishniac effects). Recent discussions of this problem in 
publications cited above use the popular spectral description 
of expected small scale variations of $\delta T/T$. This 
spectrum cannot be, however, found with both MAP and Plank 
missions and to do this special extensive observations are 
required. Here we demonstrate that the more simple direct analysis 
of the map of CMB temperature variations provides also valuable 
information about the reionization epoch and in some respect could 
be more perspective. 

As was shown in Kaiser (1984) and Ostriker \& Vishniac 
(1986), $\delta T/T$ generated after epoch $z=z_\tau$ where 
the universe last is optically thick, $\tau_e(z_\tau)\sim$1, 
are small because the different phases of any single plane 
wave cancel each other out, and the main effect is generated 
at the epoch of reionization at $z=z_{ri}$ and/or at redshifts 
when $\tau_e(z_\tau)\sim$1. For earlier reionization when 
$z_{ri}\geq z_\tau$ and the universe is optically thick, 
$\tau_e(z_{ri})\gg$1, the $\delta T/T$ averages the peculiar 
motion across the Hubble length $c/H(z_\tau)$ at the epoch 
of last scattering (Sunyaev 1978). In the case, the damping 
of perturbations is defined by the variation of optical 
depth, $\tau_e(z)$, with redshift. For later reionization, 
$z_{ri}\leq z_\tau$, the dynamic of reionization becomes 
important and random angular variations of $z_{ri}$ can 
also be seen. 

The dumping depends upon the velocity coherent length, 
$l_v$, defined by the initial power spectrum. 
Some characteristics of density and velocity fields in the 
models with Harrison-Zel'dovich initial power spectrum and 
Bardeen et al. (1986) transfer function were found in 
Demia\'nski \& Doroshkevich (1999; 2001). 
Here we show that for such models this damping is moderate 
and an amplitude $\delta T/T\sim 10^{-5} - 3\cdot 10^{-6}$ 
can be expected in minute and subminute ranges. 

\section{Period of reionization}

The processes of reheating and reionization of intergalactic gas 
have been widely discussed during last years (see, e.g., Gnedin 
\& Ostriker 1997; Baltz, Gnedin \& Silk 1997; Haiman, Rees \& 
Loeb 1997; Tegmark et al. 1997; Abel et al. 1998; Silk \& Rees 
1998; Haehnelt, Natarajan, \& Rees 1998; Miralda-Escud\'e 
et al. 2000; Haiman, Abel \& Rees 1999; Haiman \& Loeb 1999; 
Shapiro, Iliev \& Raga 1999; more references in Haiman \& Knox 
1999) but up to now it is far from clarity. 

It is agreed that the reionization is produced by the 
first population of relatively low massive galaxies with 
$$M\geq M_{min}\sim 10^7 M_\odot~{\Omega_b\over 0.1}
{0.5\over h}\left[ 10\over1+z_{vir}\right]^{3/2}, \eqno(2.1)$$
(Haiman, Rees  \& Loeb 1997) where $z_{vir}$ is redshift of 
formation of virialized cloud. The formation of more massive 
galaxies, quasars and black holes is also discussed as at 
redshifts $z\sim$3 -- 5 both bright galaxies and quasars 
are observed (see, e.g., Steidel et al. 1998; Fan et al. 2000). 

The hard problem is a reasonable estimate of luminosity of 
objects responsible for reionization and of the efficiency 
factor of reionization, $E$, that is a ratio of ionized and 
collapsed volumes. Available now estimates of this factor 
vary from $E=114$ and up to $E=10^2[20/(1+z)]^3$ (Haiman \& 
Loeb 1999; more references in Knox, Scoccimarro \& Dodelson 
1998, and Haiman \& Knox 1999). Evidently, this factor directly 
relates to the fraction of matter, $f_m(z_{ri})$, accumulated 
by objects with $M\geq M_{min}$ at the redshift of reionization, 
$z\sim z_{ri}$, and the mass conservation shows that 
$$f_m(z_{ri})\approx 1/E(z_{ri}) \eqno(2.2)$$

Our estimates based on the statistics of high density clouds 
formation (Demia\'nski \& Doroshkevich 2001) confirm that for 
both values of $E$ cited above and for larger mass of the 
heaviest DM particles $M_{DM}\geq$1 GeV the reionization is 
quite possible at redshifts $z_{ri}\sim$40 -- 20 when objects 
with $M\geq M_{min}$ incorporate $\sim$2 -- 8\% of the matter. 
At the same redshifts amorphous irregular clouds with a moderate 
overdensity above the mean density accumulate $\sim$15 -- 40\% 
of matter. Some part of these clouds can be unstable. The latest 
fraction increases up to $\sim$80\% at redshifts $z\sim$4 -- 5 
that is consistent with results obtained with high resolution 
simulations (see, e.g., Zhang et al. 1998). The expected fraction 
of homogeneously distributed gas at the period of reionization is  
$$f_g(z_{ri})\sim 0.4 - 0.7,\eqno(2.3)$$
for $z_{ri}\sim$20 -- 40, respectively.

These numerical estimates are obtained for the CDM-like 
primordial power spectrum discussed in Sec. 3. They are very 
sensitive to the shape of small scale primordial power spectrum 
and mass of DM particles. More conservative approach to the 
problem of reionization can be found in Benson et al. (2001) 
where $z_{ri}\sim$10 is discussed. Observations of small scale 
$\delta T/T$ will significantly clarify this problem. 

The reionization and reheating take place 
at the period when only a small fraction of baryonic matter is 
accumulated by high density clouds. As was shown in Peebles \& 
Juszkiewicz (1998), these clouds do not distort the expected 
variations of CMB temperature. Non the less, the reheating of 
baryonic component suppresses the small scale baryonic 
perturbation in scales lesser then the Jeans scale, $R_J$,
$$R_J\approx {a_s(1+z_{ri})\over H(z_{ri})}\approx {0.12 
\sqrt{T_4}\over\sqrt{\Omega_m z_{ri} }}h^{-1}{\rm Mpc},	\eqno(2.4)$$  
where $a_s$, $T_4 = T/10^4 K\sim 1$ and $\Omega_m$ are the sound 
speed, the expected temperature of reionized gas and the mean 
density of the universe. It essentially restricts the expected 
variations of the CMB temperature in small scale and, in fact, 
leads to the cutoff in the power spectrum of the CMB variations 
at $l\sim 3\cdot 10^4 - 10^5$ (Sec. 6). 
 
\section{Basic cosmological model and characteristics of 
the density and velocity fields}

\subsection{Parameters of cosmological model}

As a basic model, we use the spatially flat $\Lambda$CDM model 
(Jenkins et al. 1998) with parameters
$$\Omega_m=0.3,~~ \Omega_b=0.03,~~ h=0.7, ~~
\sigma_v(0)=650 km/s,				\eqno(3.1)$$						
where $\Omega_m ~\&~ \Omega_b$ are dimensionless density of 
dark matter (DM) and baryonic components, $\Omega_\Lambda=1-
\Omega_m$, $h=H_0/100km/s/$Mpc is dimensionless Hubble constant, 
and $\sigma_v(0)$ is a velocity dispersion at redshift $z=0$. 
The analysis of this simulation (Jenkins et al. 1998; Demia\'nski 
et al. 2001) shown that it reproduces well characteristics 
of large scale matter distribution observed at small redshifts 
and, so, can be used as a basic model for further investigations. 

Here we are interested in larger redshifts, $z\geq$10, where 
the influence of $\Omega_\Lambda$ becomes negligible and the 
main relations are simplified. For such redshifts we have  
$$H(z)\approx H_0\sqrt{\Omega_m}(1+z)^{3/2},\eqno(3.2)$$
$${\sigma_v(z)\over\sqrt{3} c}={1.6\cdot 10^{-3}\over 
\sqrt{z}}\sqrt{0.3\over\Omega_m}{\sigma_v(0)\over 650 km/s}=
{ w_0\over \sqrt{z}}.$$
The velocities of baryonic component suppressed at $z\sim$1000 
are practically restored already at $z\sim$50. 

Distances along a line of sight are measured by the conformal time 
and linked to the redshifts as follows:
$$\eta(z)\approx \eta_0[f_\Lambda-(1+z)^{-1/2}],\eqno(3.3)$$ 
$$\eta_0={2c\over H_0\sqrt{\Omega_m}}\approx 1.1\cdot 10^4
\sqrt{0.3\over\Omega_m}h^{-1}{\rm Mpc},$$
$$f_\Lambda\approx {3\over\sqrt{\pi}}\left({\Omega_m\over
\Omega_\Lambda}\right)^{1/6} = 1.47\left({\Omega_m\over 0.3}
{0.7\over\Omega_\Lambda}\right)^{1/6}$$

\subsection{Characteristics of density and velocity fields}

To estimate the efficiency of the formation $\delta T/T$ we 
use two characteristics of peculiar velocity field, namely, 
the coherence length, $l_v$, and the correlation functions of 
velocities. For the 
Harrison-Zel'dovich initial power spectrum of perturbations, 
$p(k)\propto k$, and for the Bardeen et al. (1986) transfer 
function, $T^{2}(x)$, the velocity coherent length was 
introduced in Demia\'nski \& Doroshkevich (1999) as follows:
$$l_v^{-2}=\int_0^\infty kT^2(k/k_0)dk = m_{-2}k_0^2,\quad
k_0={\Gamma h\over 1{\rm Mpc}}, $$
$$~~\Gamma =\Omega_mh\sqrt{
1.68 \rho_\gamma\over \rho_{rel} },\quad m_{-2} = \int_0^\infty
xT^2(x)dx\approx 0.023,$$
$$l_v\approx {6.6\over\Gamma}~\sqrt{0.023\over m_{-2}}~
h^{-1}{\rm Mpc} = 33h^{-1}{\rm Mpc}\left({0.2\over\Gamma}
\right),				       \eqno(3.4)$$
where $\rho_\gamma ~\&~\rho_{rel}$ are the density of CMB 
and relativistic particles (photons, neutrinos etc.). 

\subsubsection{Doppler effect}

The normalized velocity correlation function can be approximated 
by an expression (Demia\'nski \& Doroshkevich 1999):
$${<v_i({\bf r}_1)v_j({\bf r}_2)>\over\sigma_v^2(0)} =  
F_1(x)\delta_{ij}+x_ix_jF_2(x),			\eqno(3.5)$$
$${\bf x}={{\bf r}_1-{\bf r}_2\over l_v},\quad x=|{\bf x}|,\quad 
\sigma_s^2 = {1\over 2\pi^2}\int_0^\infty dk~p(k),$$
$$F_1(x)={3\over2\pi^2\sigma_s^2}\int_0^\infty dk~p(k){j_1(kl_vx)
\over kl_vx}\approx {1\over 1+x+a x^2},$$ 
$$F_2(x)={dF_1\over xdx}= -{3\over2\pi^2\sigma_s^2}\int_0^\infty 
dk~k^2 p(k){j_2(kl_vx)\over (kl_vx)^2}$$
where ${\bf r}_1$ and ${\bf r}_2$ are unperturbed coordinates of 
points at $z=0$ and the parameter $a\approx 0.3$ can be expressed 
through other moments of the power spectrum. 

Fits of the perturbation amplitude to the CMB anisotropy (Bunn 
\& White 1997) link model parameters (3.1) with COBE data and 
we get:
$$\sigma_s\approx 13.4h^{-1}{\rm Mpc}\left({h\over 0.7}\right)
\left({\Omega_m\over 0.3}\right)^{0.2}\left({T_Q\over 15\mu 
K}\right),$$
$$ w_0\approx 0.65{H_0\sigma_s\over c\sqrt{3}}.$$
where $T_Q$ is the amplitude of quadrupole component of anisotropy.
For the projection of the correlation function 
(3.4) on two directions, $\vec\gamma_1$ and $\vec\gamma_2$, we get:
$$F(x,\mu)= \mu\xi_v(x)-(1-\mu^2){|{\bf r}_1||{\bf r}_2|
\over l_v^2}F_2(x), 				\eqno(3.6)$$ 
$$\vec\gamma_1\vec\gamma_2=\mu\approx 1-\theta^2/2,\quad 
1-\mu\ll 1.$$
Here the function
$$\xi_v(x)=F_1(x)+x^2F_2(x)\approx {1-a x^2
\over (1+x+a x^2)^2},			 	\eqno(3.7)$$
describe the velocity correlation along a line of sight 
(for $\mu=1$). 

To estimates of the variations of CMB temperature at the period of 
reionization we will use the point separation measured along a line 
of sight (3.3) and for the characteristic correlation angle we have 
$$\theta_v={l_v\over\eta_0f_\Lambda}\approx 7'~~{0.7\over h}
\sqrt{0.3\over\Omega_m}{1.47\over f_\Lambda}.		\eqno(3.8)$$

\subsubsection{Ostriker-Vishniac effect}

The cross correlation of velocity and density fields is described 
by the second order power spectrum of perturbation as this 
correlation vanishes for the first order perturbations. This 
correlation leads to second order correction for the Doppler 
perturbations and to the Ostriker - Vishniac effect. The 
second order spectrum of baryonic component is found in 
Appendix A. 

The normalized correlation functions, $G_1$ and $G_2$, and the 
amplitude of CMB temperature variations for the OV-effect, 
$\sigma_b$, can be written as follows:
$$G_{ij}(y)=\langle v_i({\bf r}_1)\delta({\bf r}_1) v_j
({\bf r}_2)\delta({\bf r}_2)\rangle/\sigma_b^2 =	\eqno(3.9)$$ 
$$[2G_1(y)+y^2G_2(y)]\delta_{ij}-y_iy_jG_2(y),$$ 
$${\bf y} = {{\bf r}_1-{\bf r}_2\over r_J},\quad \alpha_J = 
k_0R_J\approx 10^{-2}{h\over 0.7}\sqrt{T_4{\Omega_m\over 0.3}
{20\over z_{ri}}},$$
$$r_J\approx 0.75R_J\ln\left({1\over\alpha_J}\right)\approx  0.17
\sqrt{T_4{0.3\over \Omega_m}{20\over z_{ri}}}h^{-1}{\rm Mpc},$$
$$G_1(y)\approx {1\over (1+y^2)^{0.6}},\quad
G_2(y)\approx -{1.2\over (1+y^2)^{1.6}},$$
%%$$T_b(x)\approx {x^2\exp(-0.6\alpha_J^2x^2)\over 
%%(1+2.5x+4x^2+0.37x^{2.5})^2},\quad x=k/k_0,$$
$$\sigma_b^2=f^2_{OV}\sigma_v^2(0),~ 
f_{OV}\approx\ln\left({1\over\alpha_J}\right)\left({h\over 0.7}
\right)^2\left({\Omega_m\over 0.3}\right)^{1.2}{T_Q\over 15\mu K}.$$
Here as before ${\bf r}_1$ and ${\bf r}_2$ are unperturbed 
coordinates of points at $z=0$ and $\sigma_b$ is an amplitude 
of perturbations. The Jeans scale, $R_J$, were introduced by (2.4). 

For the projection of the correlation function 
(3.9) on two directions, $\vec\gamma_1$ and $\vec\gamma_2$, we get:
$$\gamma_i\gamma_jG_{ij} = 2\mu G_1(x)+(1-\mu^2)
{|{\bf r}_1||{\bf r}_2|\over r_0^2}G_2(x), 	\eqno(3.10)$$ 
$$\vec\gamma_1\vec\gamma_2=\mu\approx 1-\theta^2/2,\quad 
1-\mu\ll 1,$$
and for the characteristic correlation angle we have 
$$\theta_J={r_J\over\eta_0f_\Lambda}\approx 2''
\sqrt{T_4{20\over z_{ri}}}{1.47\over f_\Lambda}.\eqno(3.11)$$

\section{Variations of optical depth}

Now it is generally agreed that the reionization starts 
with formation of separate bubbles number and size of 
which increase with time. At redshifts $z=z_{ri}$ these 
bubbles merge together and the high ionization degree, 
$f_i(z_{ri})\sim$1, is achieved. The maximal optical depth 
of the universe achieved due to reionization can be written 
as follows: 
$$\tau_{max} = c\sigma_c\int^\infty_0 {n_e(x)dx\over xH(x)}=
\tau_0z_{ri}^{3/2}K_f,$$
$$\tau_0 =1.4\cdot 10^{-3}\sqrt{0.3\over\Omega_m}
{h\Omega_b\over 0.02},
						\eqno(4.1)$$
$$K_f={3\over 2}\int_0^\infty\left({x\over z_{ri}}\right)
^{3/2}f_i(x){dx\over x}\geq 1.$$
Here $c ~\&~ \sigma_c = 6.6\cdot 10^{-25}{\rm cm}^2$ are the speed 
of light and the Thompson cross section, the density of electrons 
$n_e(z)\approx 1.2\cdot 10^{-5}\Omega_b(1+z)^3{\rm cm}^{-3}$, and 
a factor $K_f$ takes proper account of dynamic of reionization. 
Thus, $K_f=1$ for the instantaneous reionization at $z=z_{ri}$ 
and $K_f\geq$1 when partial ionization is reached at larger 
redshifts and $z^{3/2}f_i(z)$ slow decreases for $z\geq z_{ri}$. 
In particular, this function integrates separate ionized bubbles 
formed before the full reionization and, so, is a random function 
of line-of-sight. 

Three models of $\delta T/T$ generation should be considered. 
For earlier reionization when $\tau_{max}\gg$1 we have
$$z_\tau=z_0=\tau_0^{-2/3}=80\left({\Omega_m\over 0.3}\right
)^{1/3}\left({0.02\over\Omega_bh}\right)^{2/3}.
						\eqno(4.2)$$
In the case, the characteristics of $\delta T/T$ depend only 
upon parameters of the basic cosmological model. For 
later reionization, when $\tau_{max}\geq$1 but $z_{ri}\leq 
z_0$ we have 
$$z_0\geq z_\tau =z_{ri}K_f^{2/3}\geq z_{ri}.	\eqno(4.3)$$
For $\tau_{max}\leq$1, the generation of $\delta T/T$ is 
driven by the dynamic of the process of reionization and final 
estimates of $\delta T/T$ become more uncertain. In particular, 
in the case random angular variations of $\tau_{max}$ can be 
observed. 

When the CMB radiation is scattered in a cloud with a proper size 
$\sim l_v$ and a correlated velocity along a line-of-sight 
$v(z)$, the temperature perturbations of CMB observed with angular 
resolution smaller than a cloud size, $\theta_{cl}\sim 10'$, are 
$$\left({\delta T\over T}\right) = \sigma_c n_e(z)l_v
v(z)/c\approx 					\eqno(4.4)$$
$$2.3\cdot 10^{-6}{l_v\over 33h^{-1}{\rm Mpc}}
\left({1+z\over 20}\right)^{3/2}
{\Omega_bh^2\over 0.015}{\sigma_v(0)\over 650km/s}.$$
This amplitude is comparable with estimates of Gruzinov \& Hu 
(1998) and with the expected small scale variations 
of the CMB temperature (Secs. 5.1, 5.2). The measurable amplitude 
of $\delta T/T\sim 10^{-5}$ can be achieved only within ionized 
bubbles with an extremal velocity $\sigma_v\sim 3
\sigma_v(0)$. Formation of such bubbles is not forbidden even 
at larger redshifts but it becomes more probable in a course 
of merging of many low massive bubbles just before the reionization. 
For later reionization, $\tau_{max}\leq$1, such bubbles could be 
observed as a random angular variation of $z_{ri}$ and 
$\tau_{max}$. 

Factors $f_i(z)$ and $K_f$ characterize the redshift variation 
of number density of the homogeneously distributed ionized gas 
and must be corrected for probable matter concentration 
within low massive clouds. Small fraction of denser clouds 
provides the reionization but essential fraction of gas can 
be accumulated within irregular clouds with various sizes and 
moderate overdensity above the mean density, $\delta\rho/
\rho\sim$5 -- 10. As was shown by Peebles \& Juszkiewicz (1998),
action of this effect cannot substantially change the expected 
optical depth, $\tau$, and parameters $z_0$ and $z_\tau$ but 
for generality we will include it in our consideration. 

Now the possible formation of such clouds with variety of 
morphology is widely discussed. For example, in simulations 
at redshift $z\sim$4 (see, e.g., Zhang et al. 1998) up to 
90 -- 95\% of gas are found to be accumulated within such 
amorphous clouds that explains the suppression of observed 
Gunn-Peterson effect. Efficiency of such concentration 
at larger redshifts depends upon the amplitude and coherent 
length of density field and, so, is sensitive to the mass 
of DM particles and the shape of initial power spectrum. 

Here we will characterize action of this effect by the 
mean fraction of homogeneously distributed gas at redshift 
of reionization, $f_g = \langle f_g(z)\rangle\leq$1. 
The expected values of $f_g$ can vary in wide range, 
0.05$\leq f_g\leq$1 with redshift, coherent length $l_\rho$ 
and dynamic of reionization. This factor can be incorporated 
in relations (3.1) -- (3.4) by the substitution of $f_g\Omega_b$ 
in place of $\Omega_b$ that increases our previous estimates 
of $z_0$ and decreases estimates of $\tau_{max}$. The redshift 
variations of $f_g(z)$ can be incorporated in the factor $K_f$ 
by the substitution of $f_g(z)f_i(z)/\langle f_g(z)\rangle$ 
in place of $f_i(z)$. 

\section{Variations of CMB temperature }

The perturbations of CMB temperature caused by Doppler 
effect are defined as follows:
$$\left({\delta T\over T}\right)_D = \int_0^\infty 
{v(z)\over c}e^{-\tau_e(z)}d\tau_e(z).\eqno(5.1)$$
(see, e.g., Hu, Scott \& Silk 1994; Peebles \& Juszkiewicz 
1998). Here $v(z)$ is a line-of-sight peculiar velocity and, 
as before, $\tau_e(z)$ is the optical depth due to Thompson 
scattering of CMB.  For a cloud situated at redshift $z$ 
with a proper size 
$$D_{cl} = cH^{-1}(z)\Delta z,~~\Delta z=H(z)D_{cl}/c\ll 1,
						\eqno(5.2)$$ 
this expression becomes identical to (4.4). In general case 
it defines a random function $\delta T/T$. For Gaussian initial 
perturbations the probability distribution function of 
$\delta T/T$ is also Gaussian and its properties are 
characterized by the variance, $\sigma_D^2$: 
$$\sigma_D^2 = \int_0^\infty d\tau_e(z_1)d\tau_e(z_2){\langle 
v(z_1)v(z_2)\rangle\over c^2} e^{-\tau_e(z_1)-\tau_e(z_2)}
						\eqno(5.3)$$
$$=\int_0^\infty d\tau_e(z_1)d\tau_e(z_2){\sigma_v(z_1)
\sigma_v(z_2)\over 3 c^2} \xi_v(z_1,z_2) e^{-\tau_e(z_1)-
\tau_e(z_2)}.$$
Here the function $\xi_v(z_1,z_2)$ describes the correlation 
of peculiar velocities along a line-of-sight and, therefore, 
the expected damping of perturbations at the period of 
reionization and at smaller redshifts. This expression can 
also be obtained directly from Eqs. (25)--(32) of Hu, Scott 
\& Silk (1994). 

\subsection{Amplitude of CMB fluctuations generated 
by the Doppler effect}

For the basic model, the correlation of velocities along a 
line-of-sight are described by (3.5) with $\mu=1$ and 
$$x = {1\over\kappa_D}\left|\sqrt{z_{ri}\over z_1}-
\sqrt{z_{ri}\over z_2}\right|,$$
$$\kappa_D\approx {l_v\sqrt{z_{ri}}\over\eta_0}
\approx 1.35\cdot 10^{-2}\sqrt{{z_{ri}\over 20}{0.3\over
\Omega_m}}{0.7\over h}.				\eqno(5.4)$$

For $\kappa_D\ll$1 the expression (5.3) can be rewritten more 
explicitly as follows:
$$\sigma_D^2\approx {9\over 4}\tau_0^{2/3}w_0^2
\int_0^\infty dx_1 dx_2\xi(z_\tau x_1,z_\tau x_2)\exp[
-\tau_0(x_1^{3/2}+x_2^{3/2})]$$
$$ = 2.25\tau_0^{2/3}w_0^2\kappa_D 
K_D,\quad \tau_{max}\gg 1,			\eqno(5.5a)$$
$$\sigma_D^2\approx {9\over 4}\tau_0^2w_0^2 z_{ri}^2
\int_0^1 dx_1 dx_2\xi(z_{ri} x_1,z_{ri} x_2)e^{-\tau(z_1)-
\tau(z_2)}$$
$$ = 2.25\tau_0^2 w_0^2 z_{ri}^2\kappa_D K_D,\quad 
\tau_{max}\leq 1.				\eqno(5.5b)$$
Here $\kappa_D$ describes the main damping of perturbations 
while the factor $K_D\leq$1 takes proper account of damping and 
weights it along a line-of-sight. $K_D$ weakly depends upon small 
redshifts, slowly decreases with $\kappa_D$ and $K_D\approx$0.25 
-- 0.5 for the range of interest, $\kappa_D\sim$0.01 -- 0.1. For 
$\kappa_D\leq$0.01 $K_D$ decreases $\propto \kappa_D\ln(1/
\kappa_D)$. Parameters $ w_0$, and $\tau_0$ were introduced 
by (3.2) and (4.1). 

For the hypothetical case of earlier reionization, $\tau_{max}\gg 
1, ~~z_\tau = z_0$, we have 
$$\sigma_D\approx 3\cdot 10^{-5}\left({f_g\Omega_b\over 
0.03}\right)^{1/6}\left({0.15\over\Omega_mh^2}\right)^{5/6}
{\sigma_v(0)\over 650 km/s},			\eqno(5.9)$$
This result confirms that for so earlier reionization, 
$z_{ri}\geq z_\tau\sim 80(f_g\Omega_bh/0.02)^{-2/3}(\Omega_m
/0.3)^{1/3}, ~\tau_{max}\gg$1 the measurable variations of 
$\delta T/T\sim 10^{-5}$ can be expected with a high 
probability even if the velocities of baryonic component are 
not yet restored and $\sigma_v(0)\leq$650km/s. In the case, 
the amplitude depends mainly upon basic parameters of 
cosmological model and random angular variations of $z_{ri}$ 
caused by the random spatial distribution of earlier ionized 
bubbles are not seen.

For more realistic case of later reionization we have  
$$\sigma_D\approx 2.\cdot 10^{-3}w_0{f_g\Omega_b\over 0.03}\left(
{z_{ri}\over 20}{0.3\over\Omega_m}{0.7\over h}\right)^{3/4}
K_f^{5/6},				  \eqno(5.10)$$
where $K_f$ is given by (4.1), and $f_g\leq$1 is the mean fraction 
of homogeneously distributed gas at $z=z_{ri}$. The random factor 
$K_f$ characterizes the dynamic of reionization along a 
line-of-sight. 

The measurable variations of $\delta T/T\sim 10^{-5}$ can 
be expected with a reasonable probability for $f_g\Omega_b
z_{ri}K_f\geq$0.5 when at least higher peaks in Gaussian 
field can be really observed. The largest clouds ionized 
at higher redshifts magnify $K_f$ and can be seen 
as random higher amplitude patches. 

\subsection{Amplitude of CMB fluctuations generated 
by the Ostriker-Vishniac effect}

Amplitude of the CMB fluctuations generated by the 
Ostriker-Vishniac effect can be found in the same manner:
$$\sigma_{OV}^2 = \sigma_{D}^2
{\kappa_JK_Jf^2_{OV}\over \kappa_DK_Dz_{ri}^2},\eqno(5.11)$$
$$\kappa_JK_J=2\int_0^{z_{ri}}{dz_1dz_2 G_1(z_1,z_2)\over 
(1+z_1)(1+z_2)}e^{-\tau(z_1)-\tau(z_2)}$$
$$\kappa_J = {r_J\sqrt{z_{ri}}\over\eta_0} = 0.7\cdot 10^{-4}
\sqrt{T_4{z_{ri}\over 20}},\quad K_J\approx 53.$$
$$\sigma_{OV}\approx 0.2\sigma_{D}\left({h\over 0.7}{\Omega_m\over 
0.3}\right)^{1.5}\left({20\over z_{ri}}\right)
T_4^{1/4}{w_0\over 1.7\cdot 10^{-3} }$$
Here the function $G_1(z_1,z_2)$ describes the correlation 
of perturbations along a line-of-sight, parameters $\eta_0, r_J, 
f_{OV}, \kappa_D, K_D \& \sigma_D$ were introduced in (3.3), 
(3.9), (5.4), (5.5) and (5.10). 

\section{Angular correlation function and power spectrum 
of CMB variations}

\subsection{Power spectra}

The power spectra of CMB variations can be found directly 
from the power spectra of perturbations (Vishniac 1987, Appendix 
B \& C):
$${l^2C_l\over 2\pi}\approx 
6.4\cdot 10^{-12}{l\over\alpha}T^2\left({l\over\alpha}
\right),					\eqno(6.1)$$
$${l^2C_l\over 2\pi}\approx 
10^{-12} \left({l\over\alpha}\right)^3T_b\left({l\over\alpha}
\right).					\eqno(6.2)$$
for Doppler and Ostriker-Vishniac effects, respectively. 
In both cases the power spectra of CMB variations reproduce well 
the transfer functions of perturbations, namely $T^2 \&T_b$. The 
transfer function of baryonic perturbations, $T_b$, is introduced  
in (A.8). As was shown in Appendix B \& C for larger $l$, 
$C_l\propto z_{ri}$ for the Doppler and $C_l\propto \sqrt{z_{ri}}$ 
for the Ostriker-Vishniac effects, that verifies the domination of 
period of reionization for both effects (Vishniac 1987). 

These power spectra are plotted in Fig. 1 together with the 
spectrum of the expected primary CMB anisotropy. The 
variations of the amplitude with cosmological parameters and 
redshifts of reionization are described by (5.10), (5.11), (B.4) 
and (C.3). 

\begin{figure}
\centering
\epsfxsize=6. cm
\epsfbox{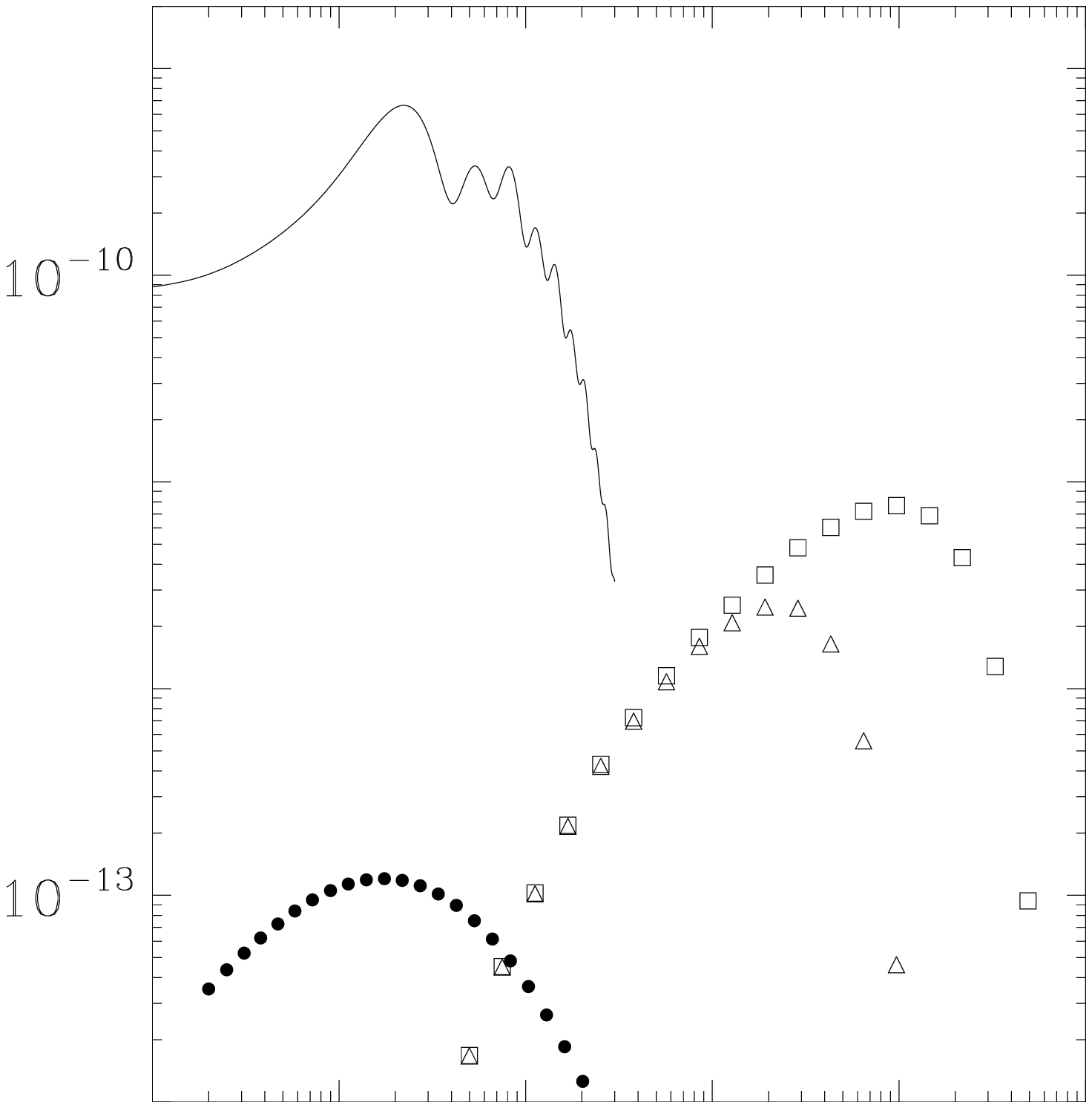}
\vspace{1.cm}
\caption{The spectrum of the CMB temperature anisotropy 
generated by the Doppler (dots) and OV-effects for 
$\alpha_J=0.05$~(triangles) ~\&~ 0.01 ~(squares) as compared 
with the expected primary CMB anisotropy (solid line). 
}
\end{figure}

\subsection{Correlation functions}

The angular correlation function of the CMB fluctuations generated 
by both effects, $w(\theta)$, can be found from the general 
expressions (3.6) and (3.9) using functions $\xi_v, F_2, G_1 ~\&~
G_2$ and the point separation  
$$x^2\approx \kappa_D^{-2}\left(\sqrt{z_{ri}\over z_1}-
\sqrt{z_{ri}\over z_2}\right)^2+$$
$$\left({\theta\over \theta_v}\right)^2\left[1-{f_\Lambda^{-1}
\over\sqrt{1+z_1}}-{f_\Lambda^{-1}\over\sqrt{1+z_2}}\right]^2,
					\eqno(6.3)$$
for the Doppler effect and similar relation for Ostriker - 
Vishniac effect.

The normalized correlation functions are plotted in Figs. 2 
\& 3 for $z_{ri}=20$ and 40. For $40\geq z_{ri}\geq 10$ these 
functions can be well fitted by expressions
$$w_(x)\approx (1+x^2)^{-\gamma},\quad x=\theta/\theta_0,
					 \eqno(6.4)$$
$$\gamma_D\approx 0.43,\quad \theta_0\approx 4.2\theta_v
\approx 30',$$
$$\gamma_J\approx 0.15,\quad \theta_0\approx 4\theta_J
\approx 8'',$$
for Doppler and Ostriker-Vishniac effects, respectively. 

\begin{figure}
\centering
\epsfxsize=6.8 cm
\epsfbox{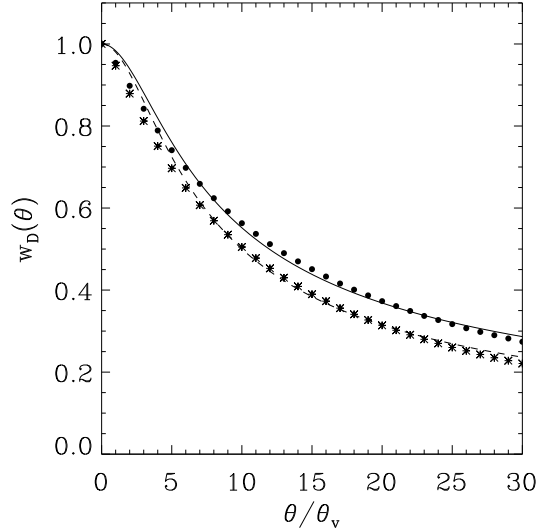}
\vspace{0.2cm}
\caption{Correlation functions of CMB fluctuations generated 
by the Doppler effect for $z_{ri}=20$ (points) and $z_{ri}=40$ 
(stars) together with fits (6.4).}
\end{figure}

\begin{figure}
\centering
\epsfxsize=6.8 cm
\epsfbox{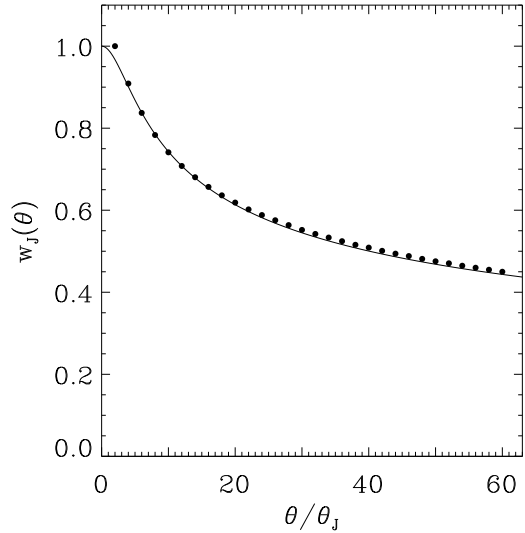}
\vspace{0.3cm}
\caption{Correlation functions of CMB fluctuations generated 
by the Ostriker-Vishniac effect for $z_{ri}=20$ together with fits 
(6.4). }
\end{figure}

\subsection{Characteristics of peaks}

The small expected amplitude of the CMB variations (5.10) and 
(5.11) indicates that the observational determination of the 
power spectrum $C_l$ is hard problem and at the first step 
only high peaks with $\delta T/T\sim 2 - 3 ~\sigma_{OV}\sim 10^{-5}$ 
could be actually observed. This means that the peak distribution 
is also an important characteristic which helps to discriminate 
the cosmological patches from the noise. 

The expected shape of peaks is described by the correlation function 
(6.4) whereas their number density is given by the Minkowski 
Functional. Application of this approach to the variations of CMB 
temperature have been discussed in Winitzki \& Kosowsky (1997), 
Schmalzing \& Gorski (1997), Novikov, Feldman \& Shandarin (1998), 
Dolgov et al. (1999). For the Gaussian distribution of the temperature 
variations with the correlation function (6.4) we get for the number 
density of positive peaks higher then threshold amplitude, $\nu=\delta 
T/T/\sigma_T$,
$$\langle n_{max}(>\nu)\rangle = {1+\gamma\over 2\pi\sqrt{3}\theta_0^2}
f(\nu),						\eqno(6.5)$$
where the function $f(\nu)$ characterizes a threshold dependence of 
$n_{max}$. This dependence helps to select and identify the peaks 
under consideration for larger statistics of the peaks with different 
$\nu$ (Novikov et al. 2001). 

The expected number density of higher Doppler peaks is: 
$$\langle n_{max}(>\nu=1)\rangle\approx (113')^{-2},\eqno(6.6a)$$
$$\langle n_{max}(>\nu=2)\rangle\approx (250')^{-2}.\eqno(6.6b)$$

The expected distribution function of CMB fluctuations generated by 
the {\it nonlinear} OV -- effect is also close (but not identical) 
to the Gaussian one. Indeed, as is well known, the distribution 
function for the product of two normal variates is 
$$f(\nu)= 1/\pi~K_0(|\nu|)\propto exp(-|\nu|),~~ |\nu|\gg 1$$
where $K_0$ is the modified Bessel function (Stuart \& Ord 1994).  
However, the integration of perturbations along a line-of-sight 
makes again the distribution of CMB fluctuations similar to the 
Gaussian one. The expected deviations from the Gaussianity are small 
for $|\nu|\leq$1 but hold more strong  for larger $|\nu|$. The degree of 
these deviations depends upon the actual number of "clouds" within 
the integral (5.1) along a line-of-sight and, therefore, upon the 
redshift of reionization. It increases for later reionization when 
small amplitude of the CMB variations is also expected. 

The application of Minkowski functional to some non-Gaussian statistics 
of CMB had been illustrated in many publications (see, e.g.,  Coles \&  
Barrow 1987, Coles 1988). Their results verify that, for example, 
deviations between the Gaussian and $\chi^2_n$ statistics with $n\geq$ 
30 do not exceed 10 -- 12\%. The significant difference between 
characteristic angles as given by (3.11) and (6.4) ($\theta_0\sim 
4\theta_J$) as well as a large difference between $\sigma_{OV}$ (5.11) 
and the possible contribution of a separate cloud with a size $\sim r_J$ 
indicate that the number of "clouds" along a line of sight is $\gg$ 20 
and the expected degree of the deviations from the Gaussianity can be 
moderate even for the highest peaks. 

Because of this we will apply expressions (6.5) for the first 
estimates of surface density of higher OV-peaks as well. We have:
$$\langle n_{max}(>\nu=1)\rangle\approx (37'')^{-2},\eqno(6.7a)$$
$$\langle n_{max}(>\nu=2)\rangle\approx (82'')^{-2}.\eqno(6.7b)$$
$$\langle n_{max}(>\nu=3)\rangle\approx (5')^{-2}.\eqno(6.7c)$$
The same expressions describe the mean number density of the negative 
peaks, $n_{min}(<\nu=-1)$, $n_{min}(<\nu=-2)$ and  $n_{min}(<\nu=-3)$. 
These estimates show that with the angular resolution $\sim 8 - 10''$ 
peaks with the amplitude $\delta T/T\sim 10^{-5}$ can be observed 
with a reasonable probability. 

\subsection{Negative intensity patches}

The discrimination of the cosmological $\delta T/T$ becomes more 
easy if we concentrate more attention on the investigation of 
{\it negative} intensity patches (Dubrovich 2000, Springel, White 
\& Hernquist 2000). The expected Gaussian distribution of 
cosmological $\delta T/T$ predicts equally probable patches 
with both positive and negative intensity, but the negative 
intensity patches cannot be produced by any background or 
foreground sources excluding only Sunyaev-Zel'dovich (SZ) 
effect in clusters of galaxies some of which can be identified 
with their x-ray radiation (at least at $z\leq$1, Refregier et 
al. 2000).

To discriminate the noise, SZ and Doppler patches their 
different frequency dependence can be used (Sunyaev \& 
Zel'dovich 1980). The Doppler shift discussed above does 
not change the spectrum of CMB and the frequency variations 
of intensity of Doppler and OV patches are 
$$\delta F_D(\chi) = F_0{\chi^4e^{-\chi}\over 
(1-e^{-\chi})^2}{\delta T\over T},\quad \chi = 
h\nu/kT_\gamma,				\eqno(6.8)$$
$$F_0 = 2.7\cdot 10^{-15} ergs(cm^2~s~sr~Hz)^{-1},\quad 
T_\gamma = 2.73 K$$
where $\chi$ is dimensionless frequency of radiation. The 
maximal intensity of these patches, $\delta F/F_0\approx 
4.9(\delta T/T)$, is achieved at $\chi\approx$3.8. It 
drops up to $\delta F/F_0\approx \delta T/T$ at $\chi\sim$1 
and $\chi\sim$8.6 . 

The frequency dependence of SZ effect (Zel'dovich \& Sunyaev 
1969; Weller, Battye \& Albrecht 1999) is 
$$\delta F_{sz}(\chi) = F_0 y_c {\chi^4e^{-\chi}\over 
(1-e^{-\chi})^2}\left[\chi{1+e^{-\chi}\over 1-e^{-\chi}}-
4\right],					\eqno(6.9)$$ 
where $y_c$ is the dimensionless Compton parameter. This 
intensity is negative at $\chi\leq$3.8 and positive at 
$\chi\geq$3.8, maximal $\delta F_{sz}/F_0=6.8y_c$ and minimal  
$\delta F_{sz}/F_0=-4.1y_c$ are achieved at $\chi=$6.5 and 
$\chi=$2.3, respectively. 

If observations are performed with a set of wavelengths in 
range 1 cm$\geq\lambda\geq$0.1 cm then the different 
frequency dependence of (6.8) and (6.9) allows to discriminate 
between patches produced by Sunyaev-Zel'dovich and Doppler 
and OV effects. 

The free-free emission of reionized plasma can, in principle, 
restrict the existence of negative intensity patches. It depends 
upon many factors such as the temperature, ionization degree 
and small scale clustering of plasma at high redshifts estimates 
of which are strongly model dependent. However, the available 
estimates of the observed intensity of the free-free emission  
$$F_{ff}\leq 3\cdot 10^{-19} ergs(cm^2~s~sr~Hz)^{-1} 
= 10^{-4}F_0					\eqno(6.8)$$
(Bartlett \& Stebbins 1991) confirm that within discussed 
range of wavelengths the cosmological patches can be observed. 
More detailed discussion of the background and foreground can be 
found in White et al. (1999), Cooray, Hu \& Tegmark (2000), and 
Refregier, Spergel \& Herbig (2000). 

\section{Summary and discussion}

Different methods of investigation of the "dark age" in the 
history of the universe evolution between $z=$1000 and $z=5$ 
were discussed during last years (see references in Sec. 1)
but most attention has been concentrated on the traditional 
spectral approach to the investigation of small scale variations 
of CMB temperature that requires extensive observations with high 
precision. The expected amplitudes of cosmological $\delta T/T$ 
obtained in Sec. 6 show that the more simple direct observation of 
higher positive and especially negative intensity patches can 
successfully complement the spectral investigations. 

The patches can be observed with already available techniques
using the aperture synthesis and/or interferometry (White et al. 
1999; Dubrovich \& Partridge 2000). Radio-frequency interferometry 
offers the advantage of virtually complete freedom from atmospheric 
emission and other systematic effects. Now such telescopes 
as the ACTA (Australia) and VLA (USA) in cm as well as IRAM 
(Plateau de Bure) and JCMT (Hawaii) by SCUBA in mm and submm 
ranges can be used to search for the patches. 

To discriminate reliably between Doppler, OV and SZ patches, 
wavelength $\lambda\leq$2 mm should be used. In future most 
promising will be ALMA, to be completed in the next decade. 
It will operate in all the atmospheric windows from 1 cm to
0.45 mm and will be able to detect sources at $\lambda\leq$0.1 
cm with the angular resolution and sensitivity of the 
present VLA at cm wavelengths.

Higher amplitudes (5.9), $\sigma_T\sim 3\cdot 10^{-5}$,  
expected for the hypothetical case of very early reionization, 
$z_{ri}\geq z_0\sim$80, depend mainly upon cosmological 
parameters, $\Omega_mh^2$ and $\sigma_v(0)$. Perhaps, so 
early reionization is possible for the tilted power spectra 
with an excess of power in small scales. In the case, 
distortions of the first generation of $\delta T/T$ can 
be essential even at angular scale $\sim 1^\circ$ (Griffiths, 
Barbosa \& Liddle 1999; Tegmark \& Zaldarriaga 2000).

More realistic scenario of later reionization, $z_{ri}\sim$40 
-- 20, predicts the existence of patches with $\sigma_T\sim  
0.5\cdot 10^{-5}$ as given by (5.10), (5.11) which can be 
observed at least as a system of the separate highest peaks. 
The amplitude is larger for earlier period of reionization due 
to growth of both $z_{ri}$ and $f_g(z_{ri})$. Reionization at 
$z_{ri}\geq$20 is possible for the tilted power spectra with 
an excess of power in small scales and/or for the Harrison-
Zel'dovich power spectrum with extremely heavy DM particles, 
$M_{DM}\sim 10^{18} - 10^{21}$keV. Existence of such particles 
is now discussed as a possible interpretation of observed high 
energy cosmic rays (Bertou, Boratav, \& Letessier-Selvon 2000; 
Nagano \& Watson 2000). In the case, possible random angular 
variations of $z_{ri}$ caused by random merging of earlier 
ionized bubbles can also be observed as the separate highest 
peaks. 

The angular distribution of the patches is driven by the 
spatial correlation of the peculiar velocity field. The same 
field is also responsible for a large scale distribution of 
galaxies and their strong concentration within filaments and 
walls (see, e.g. Demia\'nski \& Doroshkevich 1999, Demia\'nski 
et al. 2000). This concentration is observed 
in deep surveys such as the Las Campanas Redshift Survey 
(Shectman et al. 1996) and Durham/UKST redshift Survey 
(Ratcliffe et al. 1996) and is well reproduced in numerous 
simulations. For example, the mean separation of richer 
observed walls, $D_{sep}\sim 50 - 60h^{-1}$Mpc, is about 
twice of the velocity coherent length, $l_v$, as defined 
by (3.4). This implies that the correlated distribution 
of negative patches and, perhaps, the appearance of the 
{\it network of patches} with few angular minutes sizes 
can also be expected. The correlation of higher 
peaks was discussed in Heavens \& Sheth (1999).

Similar network of negative patches with the angular sizes 
$\theta\sim$1 -- $2'$ and angular separation $\theta_{sep}
\sim 5'$ was firstly observed by Carlstrom, Joy \& Grego (1996) 
with 30 GHz receiver of Owens Valley Radio Observatory in a 
course of investigations of Sunyaev-Zel'dovich effect toward 
two clusters of galaxies. 
%%Parameters of the observed network 
%%are close to our estimates obtained above for the basic  
%%cosmological model. 
The statistical reliability of these 
small amplitude negative patches is not enough, and these 
observations should be repeated with better sensitivity and 
in wider range of wavelengths, 1 cm$\geq\lambda\geq$0.1 cm. 
The frequency dependence of the amplitude and statistical 
properties discussed in Sec. 6 allows to test the attribution 
of the patches as cosmological and generated at redshifts 
$z\geq$10.

\subsection*{Acknowledgments}

This paper was supported in part by Denmark's Grundforskningsfond
through its support for an establishment of Theoretical Astrophysics
Center. It was prepared during VKD visit TAC, and he is grateful TAC 
for hospitality. We are grateful to P. Naselsky for stimulating 
conversations. We also wishes to acknowledge support from the Center 
for Cosmo-Particle Physics "Cosmion" in the framework of the project 
"Cosmoparticle Physics".

\bigskip
\centerline{\bf Appendix A}
\bigskip
\centerline{\bf Second order power spectrum}
\centerline{\bf for the Ostriker-Vishniac effect}

The cross correlation of velocity and density fields is described 
by the second order power spectrum of perturbation as this 
correlation vanishes for the first order perturbations. This 
correlation leads to second order correction for the Doppler 
perturbations and to the Vishniac - Ostriker effect. 

The second order spectrum can be written as follows:
$$\langle v_i({\bf r}_1)\delta({\bf r}_1) v_j({\bf r}_2)
\delta({\bf r}_2)\rangle =\int d^3k\Phi_{ij}(k)\exp[i{\bf k}
({\bf r}_1-{\bf r}_2)],$$
where $v_i$ and $\delta$ are peculiar velocities and relative 
density perturbations. Here 
$$\Phi_{ij}(k) = \int d^3q\langle v_i(k)\delta(k) 
v_j(q)\delta(q)\rangle = 			\eqno(A.1)$$
$$\left(\delta_{ij}-{k_ik_j\over k^2}\right)\Phi_1(k) + 
{k_ik_j\over k^2}\Phi_2(k),$$
$$\Phi_1(k) = {1\over 2}\int d^3q {p(q)\over q^2}
{p(|{\bf k}-{\bf q}|)\over ({\bf k}-{\bf q})^2}
{k^2q^2-({\bf kq})^2\over k^2q^2}
[k^2-2({\bf kq})]$$
$$\Phi_2(k) = \int {d^3q\over q^2} p(q)
p(|{\bf k}-{\bf q}|){{\bf kq}\over k^2}\left(
{k^2-({\bf kq})\over ({\bf k}-{\bf q})^2} + 
{({\bf kq})\over q^2}\right)$$

The function $\Phi_1$ describes the Vishniac - Ostriker effect 
while the function $\Phi_2$ gives the second order contribution 
to the Doppler effect. Both functions characterize the DM 
perturbations and at $k\rightarrow 0$ $\Phi_1\propto 
\Phi_2\propto k^2$ while for $k\rightarrow \infty$, $\Phi_1
\propto \Phi_2\propto k^{-3}$ and they are sensitive to the 
small scale nonlinear damping of perturbations. These functions 
are well fitted by expressions:
$$\Phi_1(x)\propto x^2(1+2.5x+4x^2+0.37x^{2.5})^{-2},
						\eqno(A.2)$$
$$\Phi_2(x)\propto x^2(1+1.45x+5.37x^2+1.25x^{2.5})^{-2}.	
						\eqno(A.3)$$
where $x=k/k_0$. The function $k^2\Phi_1(k/k_0)$ plays a role of 
the power spectra for OV-effect.

To describe perturbations of baryonic component and to take 
into account the pressure of reionized gas we will substitute  
in equations (A.1) instead of $p(k)$
$$p(k)j_0^2(kR_J),~ R_J = {a_s(1+z)\over H(z)}
\approx {0.12 \sqrt{T_4}\over\sqrt{\Omega_m z_{ri} }}h^{-1}
{\rm Mpc}.					\eqno(A.4)$$  
Here $a_s$ and $T_4 = T/10^4 K\sim 1$ are the sound speed and the 
expected temperature of reionized gas. This factor suppresses 
the small scale perturbation and instead of (A.2) we get for 
perturbations of baryonic component
$$\Phi_b(x)\approx \Phi_1(x)\exp(-0.6\alpha_J^2x^2),\eqno(A.5)$$ 
$$x=k/k_0,\quad \alpha_J=k_0R_J =10^{-2}{h\over 0.7}\sqrt{T_4
{\Omega_m\over 0.3}{20\over z_{ri}}}.$$

With this spectrum we get for the normalized spatial correlation 
functions, $G_1$ and $G_2$, the transfer function $T_b$ and 
the amplitude of OV-effect, $\sigma_b$:
$$G_{ij}(y)=\langle v_i({\bf r}_1)\delta({\bf r}_1) v_j
({\bf r}_2)\delta({\bf r}_2)\rangle/\sigma_b^2 =\eqno(A.6)$$ 
$$[2G_1(y)+y^2G_2(y)]\delta_{ij}-y_iy_jG_2(y),~ y = {r_{12}
\over r_J}={|{\bf r}_1-{\bf r}_2|\over r_J},$$ 
$$r_J\approx 0.75 R_J\ln\left({1\over\alpha_J}\right)\approx 0.17
\sqrt{T_4{0.3\over \Omega_m}{20\over z_{ri}}}h^{-1}{\rm Mpc}$$
$$G_1(y)={3\over m_b}\int_0^\infty dx x^2 T_b(x)
{j_1(kr_{12})\over kr_{12}}\approx {1\over (1+y^2)^{0.6}}$$
$$G_2(y)=-{3\over m_b}\int_0^\infty dx x^4 T_b(x)
{j_2(kr_{12})\over (kr_{12})^2}\approx -{1.2\over (1+y^2)^{1.6}}$$
$$T_b(x)\approx {x^2\exp(-0.6\alpha_J^2x^2)\over 
(1+2.5x+4x^2+0.37x^{2.5})^2},\quad x=k/k_0,\eqno(A.7)$$
$$m_b = \int_0^\infty dx x^2 T_b\approx 0.116\ln^2(\alpha_J),
						\eqno(A.8)$$
$$\sigma_b^2= \int_0^\infty d^3k \Phi_b(k) = f^2_{OV}\sigma_s^2,
						\eqno(A.9)$$
$$f_{OV}={3.7\cdot 10^{-2} \sigma_s^2\sqrt{m_b}\over l_v^2m_{-2}^{3/2} }
\approx\ln\left({1\over\alpha_J}\right)\left({h\over 0.7}\right)^2\left(
{\Omega_m\over 0.3}\right)^{1.2}{T_Q\over 15\mu K}.$$
Here as before ${\bf r}_1$ and ${\bf r}_2$ are unperturbed coordinates 
of points at $z=0$ and $\sigma_b$ is an amplitude of perturbations. 

\bigskip
\centerline{\bf Appendix B}
\bigskip
\centerline{\bf Power spectrum of CMB}
\centerline{\bf generated by Doppler effect after reionization.}
\bigskip

To obtain the power spectrum of CMB perturbations generated by 
the Doppler effect we use the general expression for the angular 
correlation function 
$$C(\mu) = z_{ri}^{-2}\int_0^{z_{ri}} 
dz_1dz_2\left(\mu\xi_v-(1-\mu^2){\eta_1\eta_2\over l_v^2}
F_2\right),\eqno(B.1)$$
where $\eta_1=\eta(z_1),\eta_2=\eta(z_2)$ and structure functions 
$\xi_v$ and $F_2$ are expressed through the power spectrum of 
primordial perturbations: 
$$\xi_v={3\over m_{-2}}\int_0^\infty xT^2(x)dx [j_0(\eta)-
2j_1(\eta)/\eta]$$
$$F_2={3\over m_{-2}^2}\int_0^\infty x^3T^2(x)dx j_2(\eta)/\eta^2,
						\eqno(B.2)$$
$$\eta = |{\vec\eta(z_1)}-{\vec\eta(z_2)}|,\quad
\eta(z)\approx\alpha x\left(f_\Lambda-{1\over\sqrt{1+z}}\right),$$
%%\quad \eta_2\approx\alpha x\left(f_\Lambda-{1\over\sqrt{1+z_2}}\right),$$
$$x=k/k_0,\quad \alpha={2ck_0\over H_0\sqrt{\Omega_m}}\approx 2\cdot 10^3{h
\over 0.7}\sqrt{\Omega_m\over 0.3},$$
and $T^2(x)$ is the transfer function of primordial perturbations.

With the 'summation theorem' (Gradshteyn \& Ryzhik 1994, 8.533), 
we get for the power spectrum of CMB perturbations:
$$C_l={3(l+1/2)\over z_{ri}^2 m_{-2}}\int_0^\infty dx 
 xT^2(x)f_l^2(x),\eqno(B.3)$$
$$f_l=\int_0^{z_{ri}} dz e^{-\tau(z)}{dj_l(\eta)\over d\eta},
\quad \eta=\eta(z)$$
%%\quad y=\alpha x\left[f_\Lambda-{1\over\sqrt{1+z}}\right].$$
This expression is identical to that obtained by Vishniac (1987).  
It can be applied to any primordial power spectrum and 
any history of reionization and indicates that $C_l$ depends upon 
both these factors. We can consider $f_l=exp[-\tau(z)]dz/d\eta$ as a 
slowly varying function and rewrite (B.3) as 
$$C_l\approx {12z_{ri} l\over m_{-2}\alpha^2}\int_0^\infty 
{dx\over x} T^2\left({x\over\alpha}\right)j_l^2(x)    
\approx {6z_{ri}\over m_{-2}\alpha^2} {1\over l}T^2\left({l\over
\alpha}\right),$$
$${l^2C_l\over 2\pi}\approx {3\over\pi}{z_{ri}\over \alpha m_{-2}}
~{l\over\alpha}T^2\left({l\over\alpha}\right)$$
$$\approx 0.5{l\over\alpha}T^2\left({l\over\alpha}\right)
{z_{ri}\over 20}{0.7\over h}\sqrt{0.3\over\Omega_m}.\eqno(B.4)$$
%%~{l\over\alpha}T^2\left({l\over\alpha}\right).$$

\bigskip
\centerline{\bf Appendix C}
\bigskip
\centerline{\bf Power spectrum of CMB generated by }
\centerline{\bf the Ostriker-Vishniac effect after 
reionization.}
\bigskip

To obtain the power spectrum of CMB perturbations generated by 
the Ostriker-Vishniac effect we use the general expression for 
the normalized angular correlation function 
$$C=\int_0^{z_{ri}}{dz_1\over 1+z_1}{dz_2\over 1+z_2}
\left[\mu G_1-(1-\mu^2){\eta_1\eta_2\over r_J^2}
G_2\right], \eqno(C.1)$$
where $\eta_1=\eta(z_1),\eta_2=\eta(z_2)$ and structure functions 
$G_1$ and $G_2$ are expressed through 
the power spectrum of primordial perturbations (Appendix A)
With the 'summation theorem' (Gradshteyn \& Ryzhik 1994, 8.533), 
we get for the power spectrum of CMB perturbations:
$$C_l={6 l^3\over m_b}\int_0^\infty dx x^2T_b(x)g_l^2(x),
						\eqno(C.2)$$
$$g_l=\int_0^{z_{ri}} {dz\over 1+z} e^{-\tau(z)}{j_l(\eta)
\over\eta}.$$
This expression is identical to that obtained by Vishniac (1987).  
The analytical expressions for $C_l$ can be found for $T_b(x)
\propto x^0, x^{-2}, x^{-4},$ ... . They show that the approximate 
general expression for $C_l$ can be written as follows:
$${l^2C_l\over 2\pi}\approx {18\over\pi}{\sqrt{z_{ri}}
\over m_b}\left({l\over\alpha}\right)^3T_b\left({l\over\alpha}
\right)\approx \eqno(C.3)$$
$$7\left({l\over\alpha}\right)^3T_b\left({l\over\alpha}
\right)\sqrt{z_ri\over 20}\left({4.6\over\ln\alpha_J}\right)^2.$$
Numerical factor $\sim$1.5 in (C.3) is found by numerical 
integration of (C.2). 
%%This relation shows that for $l< \alpha$ the function $l^2C_l$ 
%%increases $\propto l$ and it decreases $\propto l^{-3}$ for 
%%$l>\alpha$ 
%%that is similar to numerical results (Springel, White, \& 
%%Hernquist, 2001, Gnedin \& Jaffe, 2001). 


\begin{thebibliography}{}

\bibitem[1998]{Abel-et-al}
Abel~T., Anninos~P., Norman~M.L., ~\&~ Zhang~Yu., 1998, ApJ.,
 508, 518

\bibitem[1996]{Aghanim-et-al}
Aghanim~N., D\'esert~F.X., Puget~J.L., \& Gispert~R., 
1996, A\&A, 311, 1

\bibitem[1995]{Baltz-et-al}
Baltz~E.A., Gnedin~N.Y., Silk~J., 1997, ApJ., 493, L1.

\bibitem[1986]{Bardeen-et-al}
Bardeen~J.M., Bond~J.R., Kaiser~N., Szalay~A., 1986,
ApJ.,  304, 15 

\bibitem[1986]{Bartlett-steb}
Bartlett~J.G., \& Stebbins~A., 1991, ApJ., 371, 8

\bibitem[2000]{Benson-et-al}
Benson~A.J., Nusser~A., Sugiyama~N., Lacey~C.G., 2001, 
MNRAS, 320, 153

\bibitem[1986]{Bertou}
Bertou~X., Boratav~M., Letessier-Selvon~A., 2000, 
IJMP, A15, 2181

\bibitem[1986]{Bertou}
Bruscoli~M., Ferrara~A., Fabbri~R., Ciardi~B., 2000, MNRAS, 
318, 1068

\bibitem[1986]{Bunn}
Bunn~E.F. \& White~M., 1997, ApJ., 480, 6 

\bibitem[2000]{Carlstrom}
Carlstrom~J.E., Joy~M., Grego~L., 1996, ApJ., 456, L75

\bibitem[2000]{C-B}
Coles~P., Barrow~J.D., 1987, MNRAS, 228, 407

\bibitem[2000]{Cls}
Coles~P.,  1988, MNRAS, 234, 309

\bibitem[2000]{Coor}
Cooray~A., Hu~W., Tegmark~M., 2000, ApJ., 540, 1

\bibitem[1998]{DD98}
Demia\'nski~M. \& Doroshkevich~A., 1999, MNRAS,  306, 779

\bibitem[2001]{DD98}
Demia\'nski~M. \& Doroshkevich~A., 2001, MNRAS, submit.

\bibitem[1998]{DD98}
Demia\'nski~M. et al. 2001, MNRAS, submitted

\bibitem[2000]{Dolgov}
Dolgov~A. et al. 1999, IJMP D., 8, 189

\bibitem[2000]{Dubrovich}
Dubrovich~V.K., 2000, Ast.L., submitted

\bibitem[2000]{Dubrovich-part}
Dubrovich~V.K., Partridge~R.B., 2000, A\&AT, v.20, in press

\bibitem[1999]{Fan-et-al.}
Fan~X. et al., 2000, AJ., 120, 1167

\bibitem[1997]{Gnedin-Ostr}
Gnedin~N.Y., \& Ostriker~J.P., 1997, ApJ., 486, 581

\bibitem[1997]{Gnedin-Ostr}
Gnedin~N.Y., \& Jaffe~A.H., 2000, ApJ., in press, 
Astro-ph/0008469

\bibitem[1999]{Gradshteyn}
Gradshteyn~I.S. \& Ryzhik ~I.M., 1994, Table of Integrals, Series, 
and Products, Academic Press, Inc., Boston, New York, London. 

\bibitem[1999]{Griffits}
Griffiths~L.M., Barbosa~D., \& Liddle~A.R., 1999, MNRAS, 308, 854

\bibitem[1998]{Gruzinov-Hu}
Gruzinov~A., \& Hu~W., 1998, ApJ., 508, 435

\bibitem[1997]{Haiman}
Haiman~Z., Rees~M., \& Loeb~A., 1997, ApJ., 476, 458

\bibitem[1999]{Haiman}
Haiman~Z., \& Knox~L., 1999, "Microwave Foregrounds", eds. 
A. De Oliveira-Costa \& M. Tegmark (ASP, San Francisco, 1999),
p. 227

\bibitem[1998]{Haiman}
Haiman~Z., \& Loeb~A., 1999, ApJ., 519, 479

\bibitem[1998]{Haiman}
Haiman~Z., Abel~T., \& Rees~M., 2000, ApJ., 534, 11 

\bibitem[1998]{Haehnelt-et-al}
Haehnelt~M.G., Natarajan~P., Rees~M.J., 1998, MNRAS, 300, 817

\bibitem[1998]{Heavens-Sheth}
Heavens~A.F. \& Sheth~R.K., 1999, MNRAS, 310, 1062

\bibitem[1998]{Hu-silk}
Hu~W., Scott~D., Silk~J., 1994, Phys.Rev.D, 49, 648

\bibitem[1998]{Hu-silk}
Hu~W., 2000, ApJ, 529, 12

\bibitem[1998]{Hu-silk}
Hu~W., White~M., 1996, A\&A, 315, 33

\bibitem[1998]{Jaffe}
Jaffe~A.H., \& Kamionkowski~M., 1998, Phys.Rev. D, 58, 043001

\bibitem[1998]{Jenkins-et-al}
Jenkins~A. et al., 1998, ApJ., 499, 20.

\bibitem[1984]{Kaiser}
Kaiser~N., 1984, ApJ., 282, 374

\bibitem[1998]{Knox-et-al}
Knox~L., Scoccimarro~R., \& Dodelson~S., 1998, PhysRev., 81, L2004

\bibitem[1999]{Miralda-Escude}
Miralda-Escud\'e~J., Haehnelt~M., Rees~M., 2000, ApJ., 530, 1

\bibitem[1999]{Molnar-Birk}
Molnar~S.M., Birkinshaw~M., 2000, ApJ., 537, 542

\bibitem[2000]{NW}
Nagano~M., \& Watson~A.A., 2000, Rew.Mod.Phys., 72, 689

\bibitem[2000]{Nov}
Novikov~D., Feldman ~H.A. \& Shandarin~S.F., 1999, IJMP, D8, 291

\bibitem[2001]{Nov}
Novikov~D. et al., 2000, IJMP, in press

\bibitem[1986]{ostriker-vishniac}
Ostriker~J.P., \& Vishniac~E.T., 1986, ApJ., 306, L51

\bibitem[1998]{Peebles-Juz}
Peebles~P.J.E., \& Juszkiewicz~R., 1998, ApJ., 509, 483

\bibitem[1998]{Persi}
Persi~F.M., 1995, ApJ., 441,1

\bibitem[1998]{Peter}
Peterson~J.,B. et al., 1999, astro-ph/9907276

\bibitem[1996]{Ratcliffe-et-al96}
Ratcliffe,~A., Shanks,~T., Broadbent,~A., et al., 1996, MNRAS, 
281, L47

\bibitem[1999]{Rees}
Rees~M.J., 1999, 9th Annual October Astrophysics Conference 
in Maryland. Eds. S. Holt \& E. Smith. 
American Institute of Physics Press, 1999, p. 13

\bibitem[1998]{refreg}
Refregier~A., Komatsu~E., Spergel~D.N., Pen~U.L., 2000, 
Phys.Rev.D, is press, astro-ph/9912180

%%\bibitem[2000]{shaver-et-al}
%%Shaver~P.A., Windhorst~R.A., Madau~P. de Bruin~A.G., 1999, A\&A, 
%%345, 380

\bibitem[2000]{shapiro-et-al}
Shapiro~P.R., Iliev~I.T., \& Raga~A.C., 1999,  MNRAS, 307, 203

\bibitem[1996]{Shectman-et-al96}
Shectman~S.A., et al., 
%%Landy~S.D., Oemler~A., et al., 
%%Tucker~D.L., Lin~H., Kirshner~R.P., Schechter~P.L., 
1996, ApJ, 470, 172

\bibitem[1999]{Schmalzing}
Schmalzing~J. \& Gorski~K.M., 1998, MNRAS, 297, 355

\bibitem[2000]{silk-rees}
Silk~J., \& Rees~M.J., 1998, A\&A, 331, L1

\bibitem[1997]{Gnedin-Ostr}
Springel~V., White~M., \& Hernquist~L.,  2001, ApJ., in press, 
Astro-ph/0008133

\bibitem[1996]{Steidel-et-al}
Steidel~C.C., Adlerberger~K.L., Dickinson~M., Giavalisco~M.,
Pettini~M., \& Kellogg~M., 1998, ApJ.,  492, 428.

\bibitem[1994]{Stuart-ord}
Stuart~A., \& Ord~J.K., 1994, Kendall's Advanced Theory of 
Statistics, vol.1, Distribution Theory, Edward Arnold, London, 
Melburn, Auckland. 

\bibitem[1978]{Sunyaev}
Sunyaev~R.A., 1978, in Large Scale Structure of the Universe, 
ed. M.S. Longair \& J. Einasto (Dordrecht: Reidel), 393

\bibitem[1978]{Sunyaev}
Sunyaev~R.A., Zel'dovich~Ya.B., 1980, MNRAS, 190, 413

\bibitem[1994]{Tegmark-et-al}
Tegmark~ M., Silk~J., Blanchard~A., 1994, ApJ., 420, 484.

\bibitem[1997]{Tegmark-et-al}
Tegmark~ M., \& Silk~J., 1995, ApJ., 441, 458

\bibitem[1997]{Tegmark-et-al}
Tegmark~ M., Silk~J., Rees~M., Blanchard~A., Abel~T., Palla~F.,
 1997, ApJ., 474, 1.

\bibitem[2000]{Tegmark-et-al}
Tegmark~ M., \& Zaldarriaga~M., 2000, ApJ., 544, 30

%%\bibitem[1999]{Tozzi-et-al}
%%Tozzi~P., Madau~P., Meiksin~A., Rees~M.J., 2000, ApJ.,  528, 597.

\bibitem[1999]{Vishniac}
Vishniac~E.T., 1987, ApJ., 322, 597

\bibitem[1998]{Weller-et-al}
Weller~J., Battye~R.A., Albrecht~A., 1999, Phys.Rev.D, 60, 103

\bibitem[1998]{White-et-al}
White~M., Carlstrom~J.E., Dragovan~M., Holzapfel~W.L., 1999, 
ApJ., 514, 12

\bibitem[1998]{Winitzki}
Winitzki~S. \& Kosowsky~A., 1998, MNRAS, 297, 355

\bibitem[1970]{Zeldovich-Sunyaev}
Zel'dovich~Ya.B., Sunyaev~R.A., 1969, Ap\&SS,  4, 301

\bibitem[1998]{Zang et al}
Zhang~Yu., Meiksin~A., Anninos~P., Norman~M.L., 1998, ApJ.,
 495, 63

\end{thebibliography}
\end{document}